\documentclass[12pt]{article}
\usepackage{epsfig}
\newcommand{\beq}{\begin{equation}}
\newcommand{\eeq}{\end{equation}}
\newcommand{\beqa}{\begin{eqnarray}}
\newcommand{\eeqa}{\end{eqnarray}}
\newcommand{\beqar}{\begin{eqnarray*}}
\newcommand{\eeqar}{\end{eqnarray*}}

\begin{document}
\addtolength{\baselineskip}{1.5mm}

\thispagestyle{empty}

\hfill{hep-th/0204191}

\vspace{32pt}

\begin{center}
\textbf{\Large\bf Supergravity Approach to Tachyon Potential \\
 in Brane-Antibrane Systems}\\

\vspace{48pt}

Hongsu Kim\footnote{hongsu@hepth.hanyang.ac.kr}

\vspace{12pt}

\textit{Department of Physics\\ Hanyang University, Seoul,
133-791, KOREA}

\end{center}
\vspace{48pt}

\begin{abstract}

Using an exact supergravity solution representing the
$D_{p}-\bar{D}_{p}$ system, it is demonstrated that one can
construct a supergravity analogue of the tachyon potential. 
Remarkably, the (regularized) minimum value of the
potential turns out to be $V(T_{0})=-2m$ with $m$ denoting the ADM
mass of a single $D_{p}$-brane. This result, in a sense, appears
to confirm that Sen's conjecture for the tachyon condensation on
unstable $D$-branes is indeed correct although the analysis used
here is semi-classical in nature and hence should be taken with some
care. Also shown is the fact that the tachyon mass squared
$m^2_{T}$ (which has started out as being negative) can actually
become positive definite and large as the tachyon rolls down
toward the minimum of its potential. It indeed signals the possibility of
successful condensation of the tachyon since it shows that near
the minimum of its potential, tachyon can become heavy enough to
disappear from the massless spectrum. Some cosmological implications of this
tachyon potential in the context of ``rolling tachyons'' is also
discussed.

\end{abstract}

\hfill{PACS numbers : 11.25.Sq, 04.65.+e, 11.10.Kk}

\setcounter{footnote}{0}

\newpage

\section{Introduction}

Recently, it has been realized that the spectrum in some vacua of type II string theories contains not only BPS $D$-branes but
also unstable non-BPS $D$-branes \cite{non-BPS, lerda1}. And the instability of these non-BPS $D$-branes has been interpreted in the
stringy context in terms of the tachyonic mode arising in the spectrum of open strings ending on the non-BPS $D$-branes.
Indeed, there could be several ways in which one can argue why the tachyonic mode should develop in the spectrum of open strings
ending on unstable $D$-branes such as $D_{p}-\bar{D}_{p}$ systems or non-BPS $D_{p}$-branes with ``wrong'' worldvolume dimensions
in IIA/IIB theories. And perhaps one of the simplest albeit formal ways goes as follows in terms of the boundary state formalism
\cite{harvey, lerda2}.
The boundary state describing a $D_{p}$-brane can have a contribution from the closed string $NSNS$ and $RR$ sectors. And on the
side of $NSNS$ sector, there is a unique boundary state for each $p$ that implements the correct boundary conditions and survives
the corresponding GSO projection. On the side of $RR$ sector, however, there is a unique boundary state only for those
$D_{p}$-branes that can couple to a $RR$ tensor potential $A_{[p+1]}$ and for all other values of $p$, the GSO projection kills
all possible boundary states in the $RR$ sector. Namely in the boundary state formalism, a supersymmetric $RR$ charged $D_{p}$-brane
is represented by
\beq
|D_{p}>_{BPS} = {1\over \sqrt{2}}(|B>_{NSNS} \pm |B>_{RR})
\eeq
where the sign in front of the $RR$ component of the boundary state determines the $RR$ charge or equivalently the orientation of the
brane. On the other hand, for the non-supersymmetric $D_{p}$-brane with wrong values of $p$ in which case $A_{[p+1]}$ is absent and
hence carries no $RR$ charge, its boundary state contains only a $NSNS$ component
\beq
|D_{p}> = |B>_{NSNS}.
\eeq
Now for this non-supersymmetric $D_{p}$-brane with wrong $p$, the
absence of the $RR$ sector in its boundary state implies the
absence of the GSO projection in the open string loop channel and
as a result, the open string spectrum contains both the
lowest-lying tachyonic mode and the next lowest gauge field. Thus
to summarize, the spectrum of open strings ending on unstable
$D_{p}$-branes is non-supersymmetric and contains a tachyon. For
$N$-coincident $D_{p}-\bar{D}_{p}$ pairs, the worldvolume gauge
symmetry is $U(N)\times U(N)$ and the tachyon $T$ is in the
complex $(N,\bar{N})$ representation whereas for, say,
$2N$-coincident unstable $D_{p}$-branes, the worldvolume gauge
symmetry is $U(2N)$ and the tachyon is in the adjoint
represenration. And in both cases, the tachyon behaves as a Higgs
field, rolling down to the minimum of its potential and Higgsing
these gauge symmetries. Thus the determination of the associated
tachyon potential in these unstable $D_{p}$-brane systems is of
primary concern. Moreover, recently the interest in the nature of
this tachyon potential has been multipled by the idea of unstable 
brane decay via the open
string tachyon condensation on these unstable $D_{p}$-branes.
(For pioneering earlier works on tachyon condensation, see \cite{halpern}.) 
And central to this is the conjecture by Sen \cite{sen1} according to
which the tachyon potential has a minimum and the negative energy
density contribution from the tachyon potential at the minimum
should exactly cancel the sum of the tensions of the
brane-antibrane pair. Thus the endpoint of the unstable
brane-antibrane annihilation should be a closed string vacuum
without any $D$-brane. \\
In the present work, we shall attempt to
read off the supergravity analogue of tachyon potential from the
semi-classical supergravity description for brane-antibrane
interaction. To this end, we begin with the brief review of Sen's
conjecture for tachyon condensation in the brane-antibrane
annihilation process which can be stated as follows. It has been
known for sometime that a complex tachyon field $T$ with mass
squared $m^2_{T}= -1/\alpha'$ lives in the worldvolume of a
coincident $D_{2p}-\bar{D}_{2p}$ pair in type IIA string theory.
The tachyon potential should be obtained by integrating out all
other massive modes in the spectrum of open strings on the
worldvolume and it gets its maximum at $T=0$. Since there is a
$U(1)\times U(1)$ Born-Infeld gauge theory living in the
worldvolume of this brane-antibrane system, the tachyon field $T$
picks up a phase under each of these $U(1)$ gauge transformations.
As a result, the tachyon potential $V(T)$ is a function only of
$|T|$ and its minimum occurs at $T = T_{0}e^{i\theta}$ for some
arbitrary, but fixed value $T_{0}$. Now Sen goes on to argue that
at $T = T_{0}$, the sum of the tension of $D$-brane and
anti-$D$-brane and the minimum negative potential energy of the
tachyon is exactly zero
\beq
V(T_{0}) + 2M_{D} = 0 \nonumber
\eeq
with $M_{D}=T_{D}$ being the $D$-brane tension. And the philosophy
underlying this conjecture is the physical expectation that the
endpoint of the brane-antibrane annihilation would be a closed
string vacuum in which the supersymmetry is fully restored.
Unfortunately, however, the tachyon potential here $V(T)$ is not
something that can be calculated at the moment in a rigorous way
although there has been numerous attempts and considerable
achievements, say, in the context of open string field theory
\cite{osf}. In the present work, using a known, exact supergravity
solution representing a $D_{p}-\bar{D}_{p}$ system (particularly
for $p=6$), we shall demonstrate in a rigorous manner that one can
construct a ``supergravity analogue'' of the tachyon potential and
it may possesses essential nature which could be typical in
the genuine, stringy tachyon potential. Some recent works discussing 
interesting related issues on the decay of brane-antibrane systems also
appeared in the literature \cite{guijosa, koji}.

\section{Exact supergravity solution for a brane-antibrane pair}\label{ }

In this section, we shall discuss the exact supergravity solutions
representing the $D6/anti-D6$ ($D6-\bar{D6}$ heceforth) system and
the intersection of this $D6/anti-D6$ with a magnetic $RR$ flux
7-brane ($(D6-\bar{D6})||F7$ for short, henceforth). The former
possesses conical singularities which can be made to disappear by
introducing the $RR$ magnetic field (i.e., the $RR$ $F7$-brane)
and properly tuning its strength in the latter. In order to
demonstrate this, we need to along the way perform the M-theory
uplift of the $D6-\bar{D6}$ pair which leads to the Kaluza-Klein
($KK$) monopole/anti-monopole solution ($KK-dipole$ henceforth)
first discussed in the literature by Sen \cite{sen2} by embedding
the Gross-Perry \cite{gp} $KK-dipole$ solution in $D=11$
M-theory context. Indeed the $D6$-brane solution is unique among
$D_{p}$-brane solutions in IIA/IIB theories in that it is a
codimension 3 object and hence in many respects behaves like the
familiar abelian magnetic monopole in $D=4$. This, in turn,
implies that the $D6-\bar{D6}$ solution should exhibit essentially
the same generic features as those of Bonnor's magnetic dipole
solution \cite {bonnor1, emp} and its dilatonic generalizations \cite{bonnor2, emp}
in $D=4$ studied extensively in the recent literature. As we shall
see in a moment, these similarities allow us to envisage the
generic nature of instabilities common in all unstable
$D_{p}-\bar{D}_{p}$ systems in a simple and familiar manner.

\subsection{$D6-\bar{D6}$ pair in the absence of the magnetic field}

In string frame, the exact IIA supergravity solution representing
the $D6-\bar{D6}$ pair is given by \cite{youm, emp}
\beqa
ds^2_{10} &=& H^{-1/2}[-dt^2 + \sum^{6}_{i=1} dx^2_{i}] +
H^{1/2}[(\Delta+a^2\sin^2 \theta)\left({dr^2\over
\Delta}+d\theta^2\right) + \Delta \sin^2 \theta d\phi^2],
\nonumber \\
e^{2\phi} &=& H^{-3/2}, \\
A_{[1]} &=& \left[{2mra\sin^2 \theta \over {\Delta + a^2\sin^2
\theta}}\right]d\phi, ~~~F_{[2]}=dA_{[1]} \nonumber
\eeqa
where the ``modified'' harmonic function in 3-dimensional transverse space is
given by
\beq
H(r) = {\Sigma \over {\Delta + a^2\sin^2 \theta}}
\eeq
and $\Sigma = r^2-a^2\cos^2 \theta$, $\Delta = r^2-2mr-a^2$ with
$r$ being the radial coordinate in the transverse directions. The
parameter $a$ can be thought of as representing the separation
between the brane and antibrane (we will elaborate on this
shortly) and changing the sign of $a$ amounts to reversing the
orientation of the brane pair, so here we will choose, without
loss of generality, $a\geq 0$. $m$ is the ADM mass of each brane
and the ADM mass of the whole $D6-\bar{D6}$ system is $M_{ADM}=2m$
which should be obvious as it would be the sum of ADM mass of each
brane when they are well separated. It is also noteworthy that,
similarly to what happens in the Ernst solution \cite{ernst} in
$D=4$ Einstein-Maxwell theory describing a pair of
oppositely-charged black holes accelerating away from each other
(due to the Melvin magnetic universe content), this $D6-\bar{D6}$
solution in IIA theory is also static but {\it axisymmetric}
in these Boyer-Lindquist-type coordinates. As has been pointed out
by Sen \cite{sen2} in the M-theory $KK-dipole$ solution case and by
Emparan \cite{emp} in the case of generalized Bonnor's solution,
the IIA theory $D6-\bar{D6}$ solution given above represents the
configuration in which a $D6$-brane and a $\bar{D6}$-brane are
sitting on the endpoints of the dipole, i.e., $(r=r_{+},
\theta=0)$ and $(r=r_{+}, \theta=\pi)$ respectively where $r_{+}$
is the larger root of $\Delta =0$, namely
$r_{+}=m+\sqrt{m^2+a^2}$. We now elaborate on our earlier comment
that the parameter $a$ appearing in this supergravity solution can
be regarded as indicating the proper separation between the brane
and the antibrane. Notice first that for large $a$, the proper
inter-brane distance increases as $\sim 2a$. Namely,
\beqa
l &=& \int^{\pi}_{0}d\theta \sqrt{g_{\theta\theta}}|_{r=r_{+}} =
\int^{\pi}_{0}d\theta H^{1/4}(\Delta + a^2\sin^2
\theta)^{1/2}|_{r=r_{+}} \nonumber \\ &\simeq &
\int^{\pi}_{0}d\theta a\sin \theta = 2a.
\eeqa
Meanwhile, as argued by Sen \cite{sen2}, the proper inter-brane
distance vanishes when $a\rightarrow 0$. In addition, that the
limit $a\rightarrow 0$ actually amounts to the vanishing
inter-brane distance can be made more transparent as follows.
Recently, Brax, Mandal and Oz \cite{bmo} discussed the
supergravity solution representing {\it coincident}
$D_{p}-\bar{D}_{p}$ pairs in type II theories and studied its
instability in terms of the condensation of tachyon arising in the
spectrum of open strings stretched between $D_{p}$ and
$\bar{D}_{p}$. Thus now, taking the $p=6$ case for example, we
first would like to establish the correspondence between our
solution given above representing $D6-\bar{D6}$ pair generally
separated by an arbitrary distance and theirs. For specific but
appropriate values of the parameters appearing in their solution,
$(c_{2}=1 (p>3), c_{1}=0, r_{0}=m/2)$ \cite{bmo} so as to
represent a neutral, coincident $D6-\bar{D6}$ pair, their solution
is given in Einstein frame by
\beqa
ds^2_{E} &=& \left[{1-r_{0}/\tilde{r} \over
1+r_{0}/\tilde{r}}\right]^{1/4}[-dt^2 + \sum^{6}_{i=1} dx^2_{i}] +
\left[1-{r_{0}\over \tilde{r}}\right]^{1/4}\left[1+{r_{0}\over
\tilde{r}}\right]^{15/4} [d\tilde{r}^2 + \tilde{r}^2(d\theta^2 +
\sin^2 \theta d\phi^2)], \nonumber \\ e^{\phi} &=&
\left[{1-r_{0}/\tilde{r} \over 1+r_{0}/\tilde{r}}\right]^{3/2},
~~~A_{[1]} = 0
\eeqa
which, in string frame, using
$g^{E}_{\mu\nu}=e^{-\phi/2}g^{S}_{\mu\nu}$, becomes
\beqa
ds^2 &=& \left[{1-r_{0}/\tilde{r} \over
1+r_{0}/\tilde{r}}\right][-dt^2 + \sum^{6}_{i=1} dx^2_{i}] +
\left[1-{r_{0}\over \tilde{r}}\right]\left[1+{r_{0}\over
\tilde{r}}\right]^{3} [d\tilde{r}^2 + \tilde{r}^2(d\theta^2 +
\sin^2 \theta d\phi^2)], \nonumber \\ e^{\phi} &=&
\left[{1-r_{0}/\tilde{r} \over 1+r_{0}/\tilde{r}}\right]^{3/2},
~~~A_{[1]} = 0.
\eeqa
Consider now, transforming from this {\it isotropic} coordinate
$\tilde{r}$ to the standard radial $r$ coordinate
\beq
\tilde{r} = {1\over 2}[(r-2r_{0}) + (r^2-4r_{0}r)^{1/2}] ~~~{\rm
or ~inversely} ~~~r=\tilde{r}\left(1+{r_{0}\over
\tilde{r}}\right)^2.
\eeq
Then their solution describing $(N=1)D6$ and $(\bar{N}=1)\bar{D6}$
now takes the form
\beqa
ds^2 &=& \left(1-{2m\over r}\right)^{1/2}[-dt^2 + \sum^{6}_{i=1}
dx^2_{i}] +   \left(1-{2m\over r}\right)^{-1/2}[dr^2 +
r^2\left(1-{2m\over r}\right)(d\theta^2 + \sin^2 \theta d\phi^2)],
\nonumber \\ e^{2\phi} &=& \left(1-{2m\over r}\right)^{3/2},
~~~A_{[1]} = 0.
\eeqa
Clearly, this solution indeed coincides
with the $a\rightarrow 0$ (i.e., vanishing separation) limit of
our more general $D6-\bar{D6}$ solution given in eq.(4). Actually,
this aspect also has been pointed out in a recent literature \cite {teo}. 
And this confirms our earlier proposition that $a$ indeed acts as a relevant
parameter representing the proper inter-brane distance even for
very small separation. \\
Next, we turn to the conical singularity
structure of this $D6-\bar{D6}$ solution. First observe that the
rotational Killing field $\psi^{\mu} = (\partial/\partial
\phi)^{\mu}$ possesses vanishing norm, i.e., $\psi^{\mu}\psi_{\mu}
= g_{\alpha \beta}\psi^{\alpha}\psi^{\beta} = g_{\phi\phi} = 0$ at
the locus of $r=r_{+}$ as well as along the semi-infinite lines
$\theta=0, \pi$. This implies that $r=r_{+}$ can be thought of as
a part of the symmetry axis of the solution. Namely unlike the
other familiar axisymmetric solutions, for the case of the
$D6-\bar{D6}$ solution under consideration, the endpoints of the
two semi-axes $\theta =0$ and $\theta =\pi$ do not come to join at
a common point. Instead, the axis of symmetry is completed by the
segment $r=r_{+}$. And as $\theta$ varies from $0$ to $\pi$, one
moves along the segment from $(r=r_{+}, \theta=0)$ where $D6$ is
situated to $(r=r_{+}, \theta=\pi)$ where $\bar{D6}$ is placed.
Then the natural question to be addressed is whether or not the
conical singularities arise on different portions of the symmetry
axis. This situation is very reminiscent of the conical
singularity structure in the generalized Bonnor's dipole solution
in Einstein-Maxwell-dilaton theory in $D=4$ extensively studied
by Emparan \cite{emp} recently. Thus below, we explore the nature
of possible conical singularities in this $D6-\bar{D6}$ solution
in $D=10$ type IIA theory following essentially the same avenue as
that presented in the work of Emparan \cite{emp}. Namely, consider
that ; if $C$ is the proper length of the circumference around the
symmetry axis and $R$ is its proper radius, then the occurrence of
a conical angle deficit (or excess) $\delta$ would manifest itself
if $(dC/dR)|_{R\rightarrow 0}=2\pi - \delta$. We now proceed to
evaluate this conical deficit (or excess) assuming first that the
azimuthal angle coordinate $\phi$ is identified with period
$\Delta \phi$. The conical deficit along the axes $\theta =0, \pi$
and along the segment $r=r_{+}$ are given respectively by
\beqa
\delta_{(0,\pi)} &=& 2\pi - \arrowvert {\Delta \phi
d\sqrt{g_{\phi\phi}} \over \sqrt{g_{\theta\theta}}d\theta}
\arrowvert_{\theta =0, \pi} = 2\pi - \Delta \phi, \\
\delta_{(r=r_{+})} &=& 2\pi - \arrowvert {\Delta \phi
d\sqrt{g_{\phi\phi}} \over \sqrt{g_{rr}}dr} \arrowvert_{r=r_{+}} =
2\pi - \left(1 + {m^2\over a^2}\right)^{1/2}\Delta \phi \nonumber
\eeqa
where, of course, we used the $D6-\bar{D6}$ metric solution given
in eq.(4). From eq.(11), it is now evident that one cannot eliminate the conical singularities along the semi-axes
$\theta = 0, \pi$ and along the segment $r=r_{+}$ at the same time. Indeed one has the options : \\
(i) One can remove the conical angle deficit along $\theta = 0, \pi$ by choosing $\Delta \phi = 2\pi$ at the expense of the
conical angle excess along $r=r_{+}$ which amounts to the presence of a {\it strut} providing the internal pressure to
counterbalance the combined gravitational and gauge attractions between $D6$ and $\bar{D6}$. \\
(ii) Alternatively, one can instead eliminate the conical singulatity along $r=r_{+}$ by choosing $\Delta \phi =
2\pi(1+m^2/a^2)^{-1/2}$ at the expense of the appearance of the conical angle deficit
$\delta_{(0,\pi)}=2\pi[1-\{a^2/(m^2+a^2)\}^{1/2}]$ along $\theta =0, \pi$ which implies the presence of {\it cosmic strings}
providing the tension
\beq
\tau ={\delta_{(0,\pi)}\over 8\pi} = {1\over 4}\left[1 -
\left({a^2\over m^2+a^2}\right)^{1/2}\right] \nonumber
\eeq
that pulls $D6$ and $\bar{D6}$ at the endpoints apart. \\
Normally, one might wish to take the second option in which the
pair of branes is suspended by open cosmic strings, namely $D6$
and $\bar{D6}$ are kept apart by the tensions generated by cosmic
strings against the collapse due to the gravitational and gauge
attractions. And the line $r=r_{+}$, $0<\theta <\pi$ joining $D6$
and $\bar{D6}$ is now completely non-singular. This recourse to
cosmic strings to account for the conical singularities of the
solution and to suspend the $D6-\bar{D6}$ system in an equilibrium
configuration, however, might appear as a rather {\it ad hoc}
prescription. Perhaps it would be more relevant to introduce an
external magnetic field aligned with the axis joining the brane
pair to counterbalance the combined gravitational and gauge
attractions by pulling them apart. By properly {\it tuning} the
strength of the magnetic field, the attractive inter-brane force
along the axis would be rendered to vanish. Indeed this conical
singularity structure of the $D6-\bar{D6}$ system and its cure via
the introduction of the external magnetic field of proper strength
is reminiscent of Ernst's prescription \cite{ernst} for the
elimination of conical singularities of the charged $C$-metric and
of Emparan's treatment \cite{emp} to remove the analogous conical
singularities of the Bonnor's magnetic dipole solution in
Einstein-Maxwell and Einstein-Maxwell-dilaton theories and in the
present work, we are closely following the formulation of Emparan
\cite{emp}.

\subsection{$D6-\bar{D6}$ pair in the presence of the magnetic field}

In order to introduce the external magnetic field with proper
strength to counterbalance the combined gravitational and gauge
attractions and hence to keep the $D6-\bar{D6}$ pair in an
(unstable) equilibrium configuration, we now proceed to construct
the supergravity solution representing $D6-\bar{D6}$ pair
parallely intersecting with a $RR$ $F7$-brane. This can be achieved by
first uplifting the $D6-\bar{D6}$ solution in IIA theory to the
$D=11$ $KK-dipole$ solution in M-theory discussed by Sen
\cite{sen2} and then by performing a {\it twisted} KK-reduction on
this M-theory $KK-dipole$. Thus consider carrying out the
dimensional lift of the $D6-\bar{D6}$ solution given in eq.(4) to
$D=11$ via the standard KK-ansatz
\beq
ds^2_{11} = e^{-{2\over 3}\phi}ds^2_{10} + e^{{4\over 3}\phi}(dy+A_{\mu}dx^{\mu})^2
\eeq
(with $A_{[1]}=A_{\mu}d^{\mu}$ being the 1-form magnetic $RR$
potential in eq.(4)) which yields
\beqa
ds^2_{11} &=& [-dt^2 + \sum^{6}_{i=1} dx^2_{i}] +
\Sigma\left[{dr^2\over \Delta} + d\theta^2\right] \\ &+&
{1\over\Sigma}[\Delta (dy-a\sin^2 \theta d\phi)^2 + \sin^2 \theta
\{(r^2-a^2)d\phi + ady\}^2]. \nonumber
\eeqa
This is the $KK$ monopole/anti-monopole solution in $D=11$ or the
M-theory $KK-dipole$ solution first given by Sen \cite{sen2}.
Similarly to the IIA theory $D6-\bar{D6}$ solution discussed
above, this M-theory $KK-dipole$ solution represents the
configuration in which $KK$ monopole and anti-monopole are sitting
on the endpoints of the dipole, i.e., $(r=r_{+}, \theta=0)$ and
$(r=r_{+}, \theta=\pi)$ respectively. Note that unlike the IIA
theory $D6-\bar{D6}$ solution, this M-theory $KK-dipole$ solution
is free of conical singularities provided the azimuthal angle
coordinate $\phi$ is periodically identified with the standard
period of $2\pi$. Now to get back down to $D=10$, consider
performing the non-trivial point identification \beq (y, ~\phi)
\equiv (y + 2\pi n_{1}R, ~\phi + 2\pi n_{1}RB + 2\pi n_{2}) \eeq
(with $n_{1}, ~n_{2}\in Z$) on the M-theory $KK-dipole$ solution
in eq.(14), followed by the associated skew KK-reduction along the
orbit of the Killing field \beq l = \left(\partial/\partial
y\right) + B\left(\partial/\partial \phi \right) \eeq where $B$ is
a magnetic field parameter. And this amounts to introducing the
{\it adapted} coordinate \beq \tilde{\phi} = \phi - By \eeq which
is constant along the orbits of $l$ and possesses standard period
of $2\pi$ and then proceeding with the standard KK-reduction along
the orbit of $(\partial/\partial y)$. Thus we recast the metric
solution, upon changing to this adapted coordinate, in the
standard KK-ansatz
\beqa
ds^2_{11} &=& [-dt^2 + \sum^{6}_{i=1}dx^2_{i}] + \Sigma\left[{dr^2\over \Delta} +
d\theta^2\right] + {\Delta + a^2\sin^2 \theta \over \Sigma}dy^2
\nonumber \\
&+& {2[(r^2-a^2)-\Delta]a\sin^2 \theta \over
\Sigma}dy(d\tilde{\phi} + Bdy) + {\sin^2 \theta \over
\Sigma}[(r^2-a^2)^2+\Delta a^2\sin^2 \theta](d\tilde{\phi} +
Bdy)^2 \nonumber \\
&=& e^{-{2\over 3}\phi}ds^2_{10} + e^{{4\over
3}\phi}(dy+A_{\mu}dx^{\mu})^2
\eeqa
and then read off the $10$-dimensional fields as
\beqa
ds^2_{10} &=& \Lambda^{1/2}\left\{[-dt^2 + \sum^{6}_{i=1}dx^2_{i}] +
\Sigma\left[{dr^2\over \Delta} + d\theta^2\right]\right\} +
\Lambda^{-1/2}\Delta \sin^2 \theta d\tilde{\phi}^2, \nonumber \\
e^{{4\over 3}\phi} &=& \Lambda, \\
A_{[1]} &=& \Lambda^{-1}{\sin^2\theta \over \Sigma}\left\{B[(r^2-a^2)^2+\Delta a^2\sin^2 \theta]
+ a[(r^2-a^2)-\Delta]\right\} d\tilde{\phi}, \nonumber \\
F_{[2]} &=& (\partial_{r}A_{\tilde{\phi}})dr\wedge d\tilde{\phi} +
(\partial_{\theta}A_{\tilde{\phi}})d\theta\wedge d\tilde{\phi},
~~~{\rm where} \nonumber \\
\Lambda &=& {1\over \Sigma}\left\{[\Delta+a^2\sin^2\theta]+2Ba\sin^2\theta[(r^2-a^2)-\Delta]+B^2\sin^2\theta
[(r^2-a^2)^2+\Delta a^2\sin^2\theta]\right\}. \nonumber
\eeqa
Note that this solution can be identified with a $D6-\bar{D6}$ pair
parallely intersecting with a magnetic $RR$ $F7$-brane since for
$B=0$, it reduces to the $D6-\bar{D6}$ solution in eq.(4) while
for $m=0$ and $a=0$, it reduces to a $RR$ $F7$-brane solution in
IIA theory. To see this last point explicitly, we set $m=0=a$ in
eq.(19) to get
\beqa
ds^2_{10} &=& \Lambda^{1/2}\left[-dt^2 + \sum^{6}_{i=1}dx^2_{i} +
dr^2 + r^2d\theta^2\right] + \Lambda^{-1/2}r^2\sin^2\theta^2
d\tilde{\phi}^2, \nonumber \\ 
e^{2\phi} &=& \Lambda^{3/2}, \\
A_{[1]} &=& A_{\tilde{\phi}}d\tilde{\phi} = {Br^2\sin^2 \theta
\over (1+B^2r^2\sin^2 \theta)} ~~~{\rm where ~~now} \nonumber \\
\Lambda &=& (1 + B^2r^2\sin^2 \theta). \nonumber
\eeqa
Clearly, this is a magnetic $RR$ $F7$-brane solution in type IIA
theory. Also note that generally a $D_{2p}$-brane has a direct
coupling to a $RR$ $F_{(2p+1)}$-brane in IIA theory. Thus for the
case at hand, the magnetic $D6$ and $\bar{D6}$-brane content of
the solution in eq.(19) couple directly to the magnetic $RR$
1-form potential of the $F7$-brane content extracted in eq.(20)
and as a result experience static Coulomb-type force that
eventually keeps the $D6-\bar{D6}$ pair apart against the
gravitational and gauge attractions. \\
Lastly, we see if the
conical singularities which were inevitably present in the
$D6-\bar{D6}$ seed solution can now be eliminated by the
introduction of this magnetic $F7$-brane content. To do so, notice
that in this $(D6-\bar{D6})||F7$ case,
$\psi^{\mu}\psi_{\mu}=g_{\tilde{\phi}\tilde{\phi}}=0$ has roots at
the locus of $r=r_{+}$ as well as along the semi-infinite axes
$\theta =0, \pi$. Thus we need to worry about the possible
occurrence of conical singularities both along $\theta=0, \pi$ and
at $r=r_{+}$ again. Assuming that the azimuthal angle coordinate
$\tilde{\phi}$ is identified with period $\Delta \tilde{\phi}$,
the conical deficit along the axes $\theta =0, \pi$ and along the
segment $r=r_{+}$ are given respectively by
\beqa
\delta_{(0,\pi)} &=& 2\pi - \arrowvert {\Delta \tilde{\phi} d\sqrt{g_{\tilde{\phi}\tilde{\phi}}} \over
\sqrt{g_{\theta\theta}}d\theta} \arrowvert_{\theta =0, \pi} = 2\pi - \Delta \tilde{\phi}, \\
\delta_{(r=r_{+})} &=& 2\pi - \arrowvert {\Delta \tilde{\phi} d\sqrt{g_{\tilde{\phi}\tilde{\phi}}} \over
\sqrt{g_{rr}}dr}\arrowvert_{r=r_{+}} = 2\pi - \left[{r_{+}-m \over B(r^2_{+}-a^2)+a}\right]
\Delta \tilde{\phi} \nonumber
\eeqa
where, in this time, we used the $(D6-\bar{D6})||F7$ metric solution given in eq.(19).
Therefore, by choosing $\Delta \tilde{\phi} = 2\pi$ and ``tuning'' the strength of the external magnetic field as
\beq
B = {(r_{+}-m)-a \over (r^2_{+}-a^2)} = {\sqrt{m^2+a^2}-a \over 2mr_{+}}
\eeq
one now can remove all the conical singularities. As stated
earlier, this removal of conical singularities by properly tuning
the strength of the magnetic field amounts to suspending the
$D6-\bar{D6}$ pair in an (unstable) equilibrium configuration by
introducing a force exerted by this magnetic field (i.e., the $RR$
$F7$-brane) to counterbalance the combined gravitational and gauge
attractive force. To see this in a qualitative manner \cite{sen2}, recall first that, when they are well separated,
the distance between $D6$ and $\bar{D6}$ is given roughly by $\sim 2a$ as shown in eq.(6) and in this
large-$a$ limit, the magnetic field strength given above in eq.(22) is $B\simeq m/4a^2$.
Next, since both the gravitational and $RR$ gauge attractive forces between the branes would be
given by $m^2/(2a)^2$ (where we used the fact that the $RR$-charge of a $D6$-brane behaves like
$q\rightarrow m$ for large inter-brane separation as discussed earlier), the total attractive force
goes like $m^2/2a^2$. Thus this combined attractive force would be counterbalanced by the repulsive
force on the magetic dipole of the $D6-\bar{D6}$ pair, $2qB \simeq 2m(m/4a^2) = m^2/2a^2$ provided 
by the properly tuned magnetic field strength $B$ of $RR$ $F7$-brane give above in eq.(22). 

\section{Supergravity analogue of tachyon potential}\label{ }

Since our strategy is to attempt to read off the supergravity
analogue of tachyon potential from the semi-classical supergravity
description for the brane-antibrane interaction, we now start with
the evaluation of the interaction energy between the brane and
antibrane and in this work, we particularly take the case of
$D6$-brane. Before we proceed, however, a cautious comment might
be relevant. That is, throughout we shall take the parameter $a$
appearing in the supergravity solution for $D6-\bar{D6}$ system as
representing the proper distance between the brane and the
antibrane. Thus one might question if this parameter $a$ can
really serve to represent the inter-brane separation all the way
from large-$a$ to $a \rightarrow 0$, i.e., even to the case of
nearly coincident $D6-\bar{D6}$ system when the tachyonic mode is
supposed to develop in the spectrum of open strings stretched
between $D6$ and $\bar{D6}$. First, recall that we already
demonstrated that for large separation, the proper inter-brane
distance grows as $l \simeq 2a$ confirming the role of the
parameter $a$ just argued. Also recall that earlier, we exhibited
the fact that our supergravity solution for the $D6-\bar{D6}$ pair
does indeed reduce, in the limit, $a\rightarrow 0$, to the
solution for the coincident $D6-\bar{D6}$ pair known in the
literature \cite{bmo}. And this last point indicates that
evidently, $a$ still acts as a relevant parameter representing the
proper inter-brane distance even for very small separation. Thus
to conclude, we may employ $a$ as a relevant parameter in the
supergravity solution representing the inter-brane distance all
the way. Now, coming back to our main interest, notice first that
in the absence of the magnetic field (i.e., the magnetic $RR$ $F7$
brane), the total energy (i.e., the ADM-mass of the system) of the
asymptotically-flat $D6-\bar{D6}$ solution given earlier is $2m$.
This is because in the limit of infinite separation $a \rightarrow
\infty$, the interaction energy between the branes should vanish
and thus the total energy should be twice the ADM-mass of a single
brane, $m$. In the presence of the magnetic field, however, the
supergravity solution is no longer asymptotically-flat, but
instead is asymptotic to the Melvin-type universe. Nevertheless,
following the treatment suggested by Hawking and Horowitz
\cite{hawking}, it is still possible to compute the total energy
of the system by taking the generalized Melvin universe as the
reference background and hence subtracting out its infinite
energy. And the result is that the regularized total energy is
still equal to $2m$. Now for finite separation, one would expect
non-vanishing interaction energy and it would be given by
\beq
E_{int} = E_{tot} - 2M_{D6} = 2m - 2Q
\eeq
where we used the fact that for an extremal (black) brane in isolation, $M_{D_{p}}=Q$. Thus now what remains to be done is to
compute the magnetic $RR$ charge $Q$ of the $D6$-brane, in the presence/absence of the content of $RR$ $F7$-brane.

\subsection{In the absence of the magnetic field}

In order to compute the $RR$ charge of a single $D6$-brane, we
certainly need the behavior of the supergravity solution for
$D6-\bar{D6}$ pair in the region very close to each of the branes
since the charge is given by the integral $Q = (1/4\pi)\int_{S^2}
F^{n}_{[2]}$, which is over a Gaussian surface having the topology
of $S^2$ and enclosing each brane. Indeed, it is possible to
demonstrate explicitly that in the $D6-\bar{D6}$ solution given
earlier in eq.(4), the metric near $(r=r_{+}, \theta=0)$
represents that of (distorted) $D6$-brane while the metric near
$(r=r_{+}, \theta=\pi)$ does that of (distorted) $\bar{D6}$-brane.
Thus in order to study the behavior of the $D6-\bar{D6}$ solution
in the region very close to the $D6$-brane, we perform the change
of coordinates from $(r, \theta)$ to $(\rho, \bar{\theta})$ given
by the transformation law \cite{sen2, emp}
\beq
r = r_{+} + {\rho \over 2}(1+\cos \bar{\theta}), ~~~\sin^2 \theta
= {\rho \over \sqrt{m^2+a^2}}(1-\cos \bar{\theta})
\eeq
where $r_{+}=m+\sqrt{m^2+a^2}$ as given earlier. Then taking
$\rho$ to be much smaller than any other length scale involved so
as to get near each pole, the $D6-\bar{D6}$ solution in eq.(4)
becomes
\beqa
ds^2_{10} &\simeq & g^{1/2}(\bar{\theta})\left({\rho\over
q}\right)^{1/2}[-dt^2 + \sum^{6}_{i=1} dx^2_{i}] + \left({q\over
\rho}\right)^{1/2}[ g^{1/2}(\bar{\theta})(d\rho^2 +
\rho^2d\bar{\theta}^2) + g^{-1/2}(\bar{\theta})\rho^2 \sin^2
\bar{\theta}d\phi^2], \nonumber \\
e^{2\phi} &\simeq & \left({\rho\over q}\right)^{3/2}
g^{3/2}(\bar{\theta}), \\ A^{m}_{[1]} &\simeq & {mr_{+}a\over
(m^2+a^2)}g^{-1}(\bar{\theta})(1-\cos \bar{\theta})d\phi \nonumber
\eeqa
where $q=mr_{+}/\sqrt{m^2+a^2}$ and $g(\bar{\theta}) =
cos^2(\bar{\theta}/2) + [a^2/(m^2+a^2)]\sin^2(\bar{\theta}/2)$.
Namely in this small-$\rho$ limit, the geometry of the solution
reduces to that of the near-horizon limit of a $D6$-brane.
However, the horizon is no longer spherically-symmetric and is
deformed due to the presence of the other brane, i.e., the
$\bar{D6}$-brane located at the other pole. To elaborate on this
point, it is noteworthy that the surface $r=r_{+}$ is still a
horizon, but instead of being spherically-symmetric, it is
elongated along the axis joining the poles in a prolate shape.
Namely, the horizon turns out to be a {\it prolate spheroid} with
the distortion factor given by $g(\bar{\theta})$. And of course,
it is further distorted by a conical defect at the poles. And
similar analysis can be carried out near the other pole at which
$\bar{D6}$ is situated. \\ We are now ready to compute the $RR$
charge of a single $D6$-brane. Using the Gauss' law, the $RR$
charge is given by
\beqa
Q &=& {1\over 4\pi}\int_{S^2}F^{m}_{[2]} = {1\over 4\pi}\int_{\partial S^2}A^{m}_{[1]} \nonumber \\
&=& {1\over 4\pi}\left[\int^{\Delta \phi}_{0}A_{\phi}d\phi|_{\bar{\theta}=\pi}
- \int^{\Delta \phi}_{0}A_{\phi}d\phi|_{\bar{\theta}=0}\right] \\
&=& {\Delta \phi \over 4\pi}[A_{\phi}(\bar{\theta}=\pi) - A_{\phi}(\bar{\theta}=0)]
= {\Delta \phi \over 2\pi}{mr_{+}\over a} \nonumber
\eeqa
where we used the fact that the boundary of $S^2$ enclosing each $D6$-brane consists of two $S^1$s with period $\Delta \phi$, one
encircling the segment $r=r_{+}$ corresponding to $\bar{\theta}=\pi$ and the other encircling the semi-infinite line $\theta=0$
(or $\pi$) corresponding to $\bar{\theta}=0$.
Particularly, if we take $\Delta \phi = 2\pi$ to remove deficit angles along $\theta = 0, \pi$, the magnetic $RR$ charge turns out
to be
\beq
Q = {mr_{+}\over a} = m\left[{m + \sqrt{m^2+a^2}}\over a\right]
\eeq
and hence the interaction energy is given by
\beqa
E_{int} = 2m - 2Q = -2m\left[{{m+\sqrt{m^2+a^2}}\over a} - 1\right]
\eeqa
which obviously is negative, reflecting the fact that the net force between $D6$ and $\bar{D6}$ is attractive as expected.
Having obtained the interaction energy of the $D6-\bar{D6}$ system, we now construct the supergravity analogue of the tachyon
potential out of it. As was mentioned earlier in the introduction, Sen \cite{sen1} conjectured that at the minimum (i.e., the true vacuum)
of the tachyon potential characterized by the true vacuum expectation value (vev) of the tachyon field $T = T_{0}$,
the total energy of a $D_{p}-\bar{D}_{p}$ system should vanish ;
\beq
E_{tot} = V(T_{0}) + 2M_{D_{p}} = 0
\eeq
where $M_{D_{p}}$ denotes the mass of a $D_{p}$-brane. Since we look for an analogue of the tachyon potential in the supegravity
description, we first try to identify the supergravity counterpart of the equation, leading to the above conjecture, with
\beq
M_{ADM} = V(T) + 2M_{D_{p}}
\eeq
where now $M_{ADM}$ and $M_{D_{p}}$ stand for the ADM mass of the whole system and that of a single brane respectively. Then it seems
natural to assume that the interaction energy between the brane and antibrane for small but finite separation, as the one we just discussed
above, should be identified with $V(T)$ since for an isolated extremal $D_{p}$-brane, $M_{D_{p}}=Q$, namely,
$V(T) = M_{ADM} - 2M_{D_{p}} = E_{tot} - 2Q = E_{int}.$  Indeed, this identification may be justified
by the fact that the semi-classical instability in the brane-antibrane system, reflected in eq.(28) which
indicates the negative interaction energy and hence the attractive force between the branes, should
be taken over, as the inter-brane distance gets smaller, by the stringy one in terms of the tachyon 
condensation.  Thus the semi-classical 
interaction energy for larger inter-brane distance and the tachyon potential for smaller separation 
should eventually be interpolated properly.  Next, since we set $V(T) = E_{int}$, we now need to relate 
the parameter $a$ in the supergravity solution representing the inter-brane separation to the tachyon 
field expectation value $|T|$. Indeed, this proper inter-brane distance parameter $a$ in our supergravity
analysis is the only parameter that can be probed and related to $|T|$. 
To this end, we note that : \\
(i) Since the tachyon field living in the worldvolume of a $D_{p}-\bar{D}_{p}$ system is complex, its potential $V(T)$ is a function
only of $|T|^2$ \cite{sen1}. As a consequence, we expect that $a$ should be a function of $|T|^2$ as well, and \\
(ii) the tachyon potential vanishes only at $T=0$ and gets its minimum at $|T| = |T_{0}|$ \cite{sen1}. \\
Therefore, these observations lead us to suggest that
\beq
|T|^2 = |T_{0}|^2 e^{-a^2/l^2_{s}}, ~~~{\rm or} ~~~a^2 =
l^2_{s}\ln{{|T_{0}|^2\over |T|^2}}
\eeq
where the string length $l^2_{s} = \alpha'$ has been inserted for
dimensional consideration since the exponent has to be
dimensionless (note that the parameter $a$ and $m$ carry the dimensions
of the length). Our proposal, the eq.(31), to relate the proper inter-brane distance $a$ to the
tachyon field expectation value $|T|$ and hence to convert the semi-classical interaction energy
into the tachyon potential, of course, shows its inevitable limitations as a supergravity approach.
Firstly, it can only accomodate the facts that for the inter-brane separation of order 
$a\sim l_{s}$, the tachyon just develops and is sitting on the false vacuum located at $|T|\simeq 0$
and then ``rolls down'' toward the true vacuum (minimum of its potential) located at $|T|=|T_{0}|$
which is assumed to happen when the brane and the antibrane precisely coincide, i.e., $a=0$. And this
is because we have in this supergravity analysis only a single free parameter $a$ (i.e., the proper
inter-brane distance) but nothing else to probe the behavior of the tachyon potential.
In a more sophisticated stringy (or we should say, string field theoretical) approach, of course, 
one would, in principle, be able to determine the behavior of tachyon potential as a function of
$|T|$ for a {\it given} inter-brane separation $a$ which lies in the interval $0\leq a \leq l_{s}$.
As anyone would admit, however, the genuine tachyon potential is not something that can be calculated
at the moment as a closed form in a rigorous manner. Indeed, it is this difficulty that led us to
consider instead the supergravity approach and attempt to read off, if possible, the supergravity analogue
of the tachyon potential. Thus we should be content with this outcome of the supergravity 
analysis but still having in mind the limitations it inevitably possesses.
Finally, plugging eq.(31) into the expression for the interaction
energy obtained above, we finally can write down the supergravity
analogue of tachyon potential as
\beq
V(T) = -2m\left[l^{-1}_{s}\left(\ln{{|T_{0}|^2\over
|T|^2}}\right)^{-1/2} \left\{m + \left(m^2 +
l^2_{s}\ln{{|T_{0}|^2\over |T|^2}} \right)^{1/2}\right\} -
1\right]
\eeq
where $m$ denotes the ADM mass of a single $D6$-brane. \\
\begin{figure}[hbt]
\centerline{\epsfig{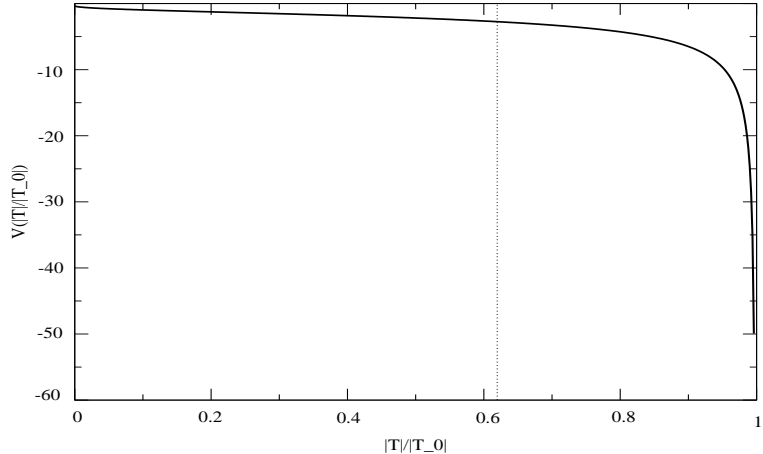}}
\caption{\small Tachyon potential in the absence of the magnetic field. The vertical line indicates the borderline 
$|T|=e^{-1/2}|T_{0}|$ within which the supergravity analysis can be safely trusted.}
\end{figure}
\\
Note that $V(T)=0$ at $T=0$ consistently with the
expectation but $V(T_{0})\rightarrow -\infty$, namely the minimum
of the tachyon potential constructed from this supergravity
description is {\it unbounded below}. This is rather discouraging
but it is interesting to note that by introducing an appropriate
magnetic field (i.e., the $RR$ $F7$-brane) aligned with the axis
joining the brane pair with the strength $B$ removing all the
conical singularities of the $D6-\bar{D6}$ solution, this tachyon
potential can be {\it regularized}, i.e., can be made to be
bounded from below. Thus now we turn to this case.

\subsection{In the presence of the magnetic field}

Similarly to the case without the magnetic field discussed above,
we can obtain the near-horizon limit of the $D6$-brane parallely
intersecting with a $RR$ $F7$-brane and located at $(r=r_{+},
\theta=0)$ by performing the change of coordinates given in eq.(24)
and taking the limit $\rho \rightarrow 0$ . In this limit the
$(D6-\bar{D6})||F7$ solution becomes
\beqa
ds^2_{10} &\simeq & g^{1/2}(\bar{\theta})\left({\rho\over
q}\right)^{1/2}[-dt^2 + \sum^{6}_{i=1} dx^2_{i}] + \left({q\over
\rho}\right)^{1/2}[ g^{1/2}(\bar{\theta})(d\rho^2 +
\rho^2d\bar{\theta}^2) + g^{-1/2}(\bar{\theta})\rho^2 \sin^2
\bar{\theta}d\phi^2], \nonumber \\
e^{2\phi} &\simeq & \left({\rho\over q}\right)^{3/2}
g^{3/2}(\bar{\theta}), \\
A^{m}_{[1]} &\simeq & q\left[{a\over \sqrt{m^2+a^2}} + 2Bq\right]
g^{-1}(\bar{\theta})(1-\cos \bar{\theta})d\phi \nonumber
\eeqa
where now $q = mr_{+}/\sqrt{m^2+a^2}$ and $g(\bar{\theta}) =
cos^2(\bar{\theta}/2) + \left[{a/\sqrt{m^2+a^2}} + 2Bq\right]^2
\sin^2(\bar{\theta}/2)$.  Again, the metric in this small-$\rho$
limit, i.e., near $(r=r_{+}, \theta=0)$ represents that of
(distorted) $D6$-brane with deformation factor given by
$g(\bar{\theta})$. Then as before, using the Gauss' law, the $RR$
charge is given by
\beqa
Q &=& {1\over 4\pi}\int_{S^2}F^{m}_{[2]} \\
&=& {\Delta \phi \over 4\pi}[A_{\phi}(\bar{\theta}=\pi) - A_{\phi}(\bar{\theta}=0)] =
{\Delta \phi \over 2\pi}q\left[{a\over \sqrt{m^2+a^2}} + 2Bq\right]^{-1}.  \nonumber
\eeqa
Particularly, for the strength $B$ of the magnetic field which eliminates all the conical singularities at once
\beq
B = {(r_{+}-m)-a \over (r^2_{+}-a^2)} = {1\over 2q}\left(1 - {a\over \sqrt{m^2+a^2}}\right)
\eeq
and with $\Delta \phi = 2\pi$, the magnetic $RR$ charge turns out to be
\beq
Q = q = {mr_{+}\over \sqrt{m^2+a^2}}
\eeq
and hence in this time the interaction energy is given by
\beqa
E_{int} = 2m - 2Q = -{2m^2 \over \sqrt{m^2+a^2}}.
\eeqa
Again, this is negative indicating the attractive nature of the force between $D6$ and $\bar{D6}$.
What remains then is to write down the supergravity analogue of the tachyon potential using our ansatz given in eq.(31) relating the
inter-brane distance parameter $a$ to the expectation value of the tachyon field $|T|$ as we did before. The result is
\beq
V(T) = -{2m^2 \over \left[m^2 + l^2_{s}\ln{{|T_{0}|^2\over |T|^2}}
\right]^{1/2}} = - 2m^2\left[m^2 - l^2_{s}\ln{{|T|^2\over
|T_{0}|^2}}\right]^{-1/2}
\eeq
where again $m$ denotes the ADM-mass of a single $D6$-brane. \\
\begin{figure}[hbt]
\centerline{\epsfig{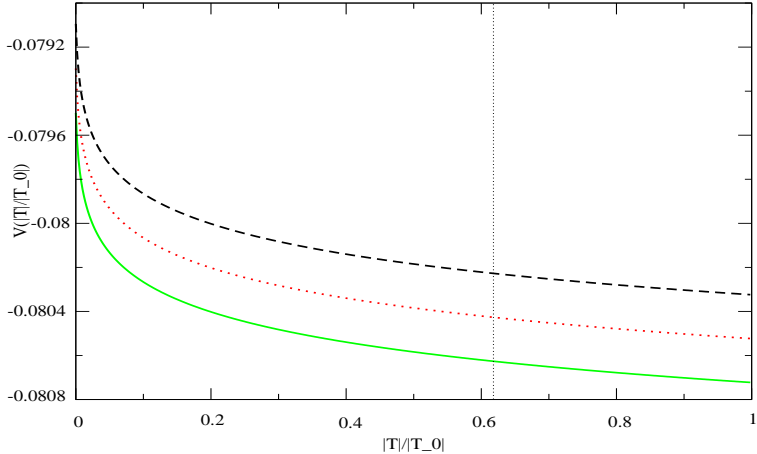}}
\caption{\small Regularized tachyon potential in the presence of the magnetic field. Three curves are shown and they indicate that
as the ratio $m^2/l^2_{s}$ gets larger, the potential gets deeper.}
\end{figure}
\\
Note that in order for the supergravity description to be valid, one should
demand in this expression for the tachyon potential, $m^2 \geq
l^2_{s}$. Nevertheless, in the following discussions, we shall
allow for $a\rightarrow 0$ (or equivalently, $|T|\rightarrow
|T_{0}|$) assuming that the supergravity analogue of the
(regularized) tachyon potential in eq.(38) would possess generic
nature which is typical in the genuine, stringy tachyon
potentials.  Again $V(T)=0$ at $T=0$ consistently with our
expectation. What is remarkable is that $V(T_{0})= - 2m$, namely
the minimum of the tachyon potential is now bounded from below.
Indeed, this regularization of the interaction energy which has
here been identified with the tachyon potential can be understood
as follows. First in the absence of the external magnetic field,
since there is combined gravitational and gauge attractions and
since the parameter $a$ can serve to represent the proper
inter-brane distance even for very small separation, the
brane-antibrane interaction energy naturally blows up as $\sim
{-1/ a}$ when the branes are brought close enough each other as we
have seen in eq.(28). In the presence of the magnetic field,
however, the strength $B$ of the magnetic field, which has been
tuned to remove all the conical singularities in the $D6-\bar{D6}$
solution, counterbalances the gravitational and gauge attractions
for large separation. As a result, even for small inter-brane
separation $a\rightarrow 0$, the introduction of the magnetic
field with properly tuned strength has an effect of diminishing
the brane-antibrane interaction energy to a finite magnitude
$V(T_{0})=-2m$ which is of particular significance, namely twice
the ADM mass of each $D6$-brane in isolation.  In addition, notice
that even when the tachyon potential gets its minimum value
$V(T_{0})= - 2m$, the total energy of this $D6-\bar{D6}$ system is
still $M_{ADM}=2m$ rather than becoming zero (which would
correspond to the supersymmetric vacuum). Indeed, the total energy
of the $D6-\bar{D6}$ system is conserved as this value for all
inter-brane distance since our analysis presented here is
semi-classical in nature based on the explicit supergravity
solution. In the stringy description, however, the total energy
increases with the inter-brane distance since the contribution to
the total energy coming from the open string stretched between the
brane and the antibrane grows from, say, $V(T_{0})=-2m$ to
$V(T=0)=0$ due to the constant tension of the string. Namely,
pulling the $D_{p}-\bar{D}_{p}$ pair farther and father apart
raises the mass of the tachyon above zero and eliminate the
perturbative instability. Therefore in the context of this stringy
description, Sen's conjecture $E_{tot} = V(T_{0}) + 2M_{D6} = 0$
does not contradict with the naive semi-classical notion of energy
conservation and in this sense, our result in the context of the
supergravity description, $V(T_{0}) + 2Q = 2m (constant)$ should
not be taken as being inconsistent with Sen's conjecture. Rather,
the failure $V(T_{0}) + 2Q \neq 0$ just reveals the limitation of
this semiclassical treatment in which the open string degree of
freedom is manifestly absent. To summarize, since $m$ denotes the
ADM mass of a single $D6$-brane, our result, $V(T_{0})=-2m$, in a
sense, confirms that, after all, Sen's conjecture was indeed
correct although in our supergravity description, the ($RR$)
charge term $2Q$, namely, the contribution to the total energy
coming from the pair of extremal branes, readjust its value (to
$2Q=4m$) to maintain the value of total energy as the conserved
value, $E_{tot} = 2m$. Namely, in this semi-clasical supergravity
description, the guiding principle is the conserved total energy,
$E_{tot}=M_{ADM}=V(T)+2Q=2m$. As the pair of branes approach each
other, the interaction energy $V(T)$ decreases starting from zero
towards its minimum $V(T_{0})=-2m$, but the brane charge (or mass)
term $2Q$ also keeps readjusting its value to compensate the
monotonously decreasing behavior of $V(T)$ and thus maintain the
total energy as the constant value $M_{ADM}=2m$. Meanwhile in the
stringy description, the guiding principle is that the brane mass
term remains unchanged since here the brane mass is really the
brane tension, $2M_{D6}=2T_{D6}$ and the total energy $E_{tot} =
V(T) + 2M_{D6}$ decreases monotonously from $2M_{D6}$ towards zero
(i.e., vacuum). And this is because, as the tachyon field rolls
down its potential, $V(T)$ decreases again monotonously from
$V(T=0)=0$ to $V(T_{0})=-2M_{D6}$. Therefore from the comparison
between these two descriptions, we may expect that the
supergravity analogue of tachyon potential obtained in this work
should possess essentially the same nature as that of genuine
stringy tachyon potential. Nevertheless, in a stricter sense, the
expressions for $V(T)$ obtained in the present work should be
regarded as a supergravity analogue of the tachyon potential
rather than as a semi-classical limit of a genuine stringy tachyon
potential.

\subsection{Other natures of the regularized tachyon potential}

Next, it is tempting to expect that the tachyon potential
constructed in the present work based on the exact supergravity
solution for a brane- antibrane system might possess generic
nature common to those in all non-BPS brane systems in type II
theories. To be a little more concrete, since other non-BPS
$D_{p}$-branes, i.e., $p=odd$ for IIA-theory and $p=even$ for
IIB-theory can be obtained from the $D_{p}-\bar{D}_{p}$ solution
via the $(-1)^{F_{L}}$ projection, one may expect that the tachyon
potentials there presumably would have essentially the same
features as those of the tachyon potential found here for the
$D6-\bar{D6}$ system. Let us elaborate on this point in some more
detail. And to do so, we need to recall the action of
$(-1)^{F_{L}}$ on the coincident $D_{p}-\bar{D}_{p}$ system, where
$F_{L}$ denotes the contribution to the {\it spacetime fermion
number} from the left-moving sector of the string worldsheet.
$(-1)^{F_{L}}$ is known to be an exact symmetry of the type
IIA/IIB string theories. To summarize its action rule, first it
has trivial action on the worldsheet fields. Next, acting on the
closed string Hilbert space, it changes the sign of all the states
on the left-moving Ramond sector, but does not change anything
else. From this rule, it follows that the spacetime fields
originating in the $RR$ sector of the worldsheet change sign under
$(-1)^{F_{L}}$. Since a $D_{p}$-brane is charged under $RR$ field
and a $\bar{D}_{p}$-brane is a $D_{p}$-brane of opposite
orientation carrying opposite $RR$ charge, it follows that
$(-1)^{F_{L}}$ takes a $D_{p}$-brane to a $\bar{D}_{p}$-brane and
vice versa. Thus a single $D_{p}$-brane or a single
$\bar{D}_{p}$-brane is not invariant under $(-1)^{F_{L}}$, but a
coincident $D_{p}-\bar{D}_{p}$ system is invariant under
$(-1)^{F_{L}}$. Hence it is relevant to study the action of
$(-1)^{F_{L}}$ on the open strings living on the worldvolume of
$D_{p}-\bar{D}_{p}$ system. We are now ready to define a non-BPS
$D_{2p}$-brane ($D_{(2p+1)}$-brane) in IIB (IIA) string theory and
it can be done by following the steps listed below \cite{sen1} ; \\ 
(i) Start with a $D_{2p}-\bar{D}_{2p}$ pair in IIA theory and take the
orbifold of this configuration by $(-1)^{F_{L}}$. \\ 
(ii) In the bulk, modding out IIA by $(-1)^{F_{L}}$ gives IIB. \\ 
(iii) Acting on the open strings living on the $D_{2p}-\bar{D}_{2p}$ system
worldvolume, $(-1)^{F_{L}}$ projection keeps states with
Chan-Paton (CP) factor $I$ and $\sigma_{1}$ and discards states
with CP factors $\sigma_{3}$ and $i\sigma_{2}$. \\ 
This defines a
non-BPS $D_{2p}$-brane of IIB string theory and in a similar
manner, starting from a $D_{(2p+1)}-\bar{D}_{(2p+1)}$  pair of IIB
theory, and modding it out by $(-1)^{F_{L}}$, one also can define
a non-BPS $D_{(2p+1)}$-brane of IIA theory. And in order to see
that these resultant non-BPS $D$-branes represent a single object
rather than a pair of objects, note that, before the projection,
the degrees of freedom of separating the two branes reside in the
sector with CP factor $\sigma_{3}$ which has been projected out by
step (iii) above. Thus these degrees of freedom of separating the
pair are lost. Now we have established the fact that type IIA
(IIB) string theory contains BPS $D$-branes of even (odd)
worldvolume dimension and non-BPS $D$-branes of odd (even)
dimensions. Then in the same spirit as in the case of coincident
$D_{p}-\bar{D}_{p}$ pair, Sen argued that at the minimum of the
tachyon potential, the sum of the tension of the non-BPS
$D_{p}$-brane, $\tilde{M}_{D_{p}}$ and the minimum (negative)
tachyon potential energy $\tilde{V}(\tilde{T}_{0})$ should exactly
be zero ; \\
\beq
\tilde{V}(\tilde{T}_{0}) + \tilde{M}_{D_{p}} = 0.
\eeq
Therefore, it may seem that these observations suggest that the
behavior of the tachyon potential $\tilde{V}(\tilde{T})$ arising
in the non-BPS brane and that of the tachyon potential arising in
the unstable $D_{p}-\bar{D}_{p}$ system would be essentially the
same. Namely, since $D_{p}-\bar{D}_{p}$ system and non-BPS
$D_{p}$-brane are related to each other via a $(-1)^{F_{L}}$
projection, the latter might inherit the generic nature of its
instability from that of the former and hence the genuine stringy
tachyon potentials $V(T_{0})$ and $\tilde{V}(\tilde{T}_{0})$
associated with the $D_{p}-\bar{D}_{p}$ system and the non-BPS
$D_{p}$-brane respectively may share some generic features. Then
next in the context of supergravity approximation like the one
discussed in the present work, one may naturally ask whether the
same is true. That is, one might be tempted to expect that in the
vanishing inter-brane distance limit $a\rightarrow 0$, the
supergravity analogue of the open string tachyon potential that
has been constructed in the present work based on the exact
supergravity solution representing a $D_{p}-\bar{D}_{p}$ pair
$(p=6)$, may well reflect generic features that would also be
typical in the tachyon potential associated with the non-BPS
$D_{p}$-brane. This naive expectation, however, should be taken
with a caution and actually it does not appear to be the case. The
rationale goes as follows. Recall that essentially we identified
the brane-antibrane interaction energy for small but finite
separation with the open string tachyon potential in the unstable
$D6-\bar{D6}$ system and to achieve this, the parameter $a$
probing the proper inter-brane distance has been translated into
the vev of the tachyon field $|T|$ as given in eq.(31). As pointed
out above, however, the non-BPS $D_{p}$-brane possesses no such
degree of freedom as the inter-brane separation. Indeed when going
from the $D_{p}-\bar{D}_{p}$ system to the non-BPS $D_{p}$-brane,
the degrees of freedom of separating the two branes which reside
in the sector with CP factor $\sigma_{3}$ have been projected out
and thus are completely lost. Therefore it is hard to imagine how
the supergravity analogue of the tachyon potential in the
$D_{p}-\bar{D}_{p}$ system whose dependence on the tachyon field
$|T|$ originates from the parameter probing the inter-brane
distance can also reflect the generic nature of the tachyon
potential in the non-BPS $D_{p}$-brane. To conclude, it seems safe
to say that the supergravity analogue of the tachyon potential
constructed in this work can only represent the generic feature of
instability in the $D_{p}-\bar{D}_{p}$ system and we need some
other approach to obtain the (possibly supergravity analogue of)
tachyon potential in the non-BPS $D_{p}$-brane. \\ Thus far, we
have discussed the constuction of supergravity analogue of tachyon
potential from the exact supergravity solution describing
$D6-\bar{D6}$ pair in IIA theory. And in doing so, we relied on
the philosophy that since the type IIA supergravity is supposed to
be a low energy effective theory of the type IIA string theory,
the analogue of tachyon potential thus obtained should reflect
essential features typical in the genuine, stringy tachyon
potential. In type IIA string theory which contains non-BPS
$D_{2p}-\bar{D}_{2p}$ pairs, tachyonic mode with mass squared
$m^2_{T}=-1/\alpha'$ is known to develop in the spectrum of open
string stretched between coincident brane-antibrane pair. Thus we
are now naturally led to the question of whether we can, from the
vanishing separation (between $D6$ and $\bar{D6}$) limit of the
supergravity analogue of the tachyon potential obtained earlier,
read off the value of the supergravity analogue of the tachyon
mass squared $m^2_{T}$ and if indeed we can, how it can actually
be compared with the stringy value known. And here, we would like
to know whether the quantity $m^2_{T}$ (which has started out as
being negative) can actually become positive definite as the
tachyon rolls down toward the minimum of its potential. If the
answer is yes, it certainly would signal the successful
condensation of the tachyon. In the following we address this
issue. We start with the regularized tachyon potential given in
eq.(38) that we have extracted from the supergravity solution
representing a $D6-\bar{D6}$ pair intersecting with a magnetic
$RR$ $F7$-brane. Since the tachyonic mode is supposed to develop
in the limit of nearly coincident brane-antibrane pair and since
the generic form of the action for the (complex) tachyon field
would look like (we work in the convention $g_{\mu\nu} =
diag(-,+,...,+)$)
\beq
S_{T} = -{T_{9}\over \sqrt{\alpha'}}\int d^{10}x\sqrt{g}[g^{\mu
\nu}\partial_{\mu}T^{*}\partial_{\nu}T + V(|T|)]
\eeq
with $T_{9}$ being the tension of $D9$-brane, we suggest the
supergravity analogue of the open string tachyon mass squared as
\beq
m^2_{T} = {d^2V(|T|)\over d|T|d|T|} = {2m^2l^4_{s}\over |T|^2}
\left[m^2-l^2_{s}\ln{{|T|^2\over |T_{0}|^2}}\right]^{-5/2}
\left\{l^{-2}_{s}\left[m^2-l^2_{s}\ln{{|T|^2\over |T_{0}|^2}}\right]-3\right\}.
\eeq
Now, we would like to evaluate the
tachyon mass squared particularly when the tachyon field is near
the minimum of its potential $|T|=|T_{0}|$. Then plugging 
$|T|=|T_{0}|$ into eq.(41) above yields
\beq
m^2_{T} = {2l^4_{s}\over m^3 |T_{0}|^2}(m^2l^{-2}_{s} - 3).
\eeq
As has been pointed out earlier, since $m$ denotes the ADM mass of
a single $D6$-brane, we may demand $ml^{-1}_{s} \geq 1$ in order
for our supergravity description to be valid. (It should be clear, from eq.(24), that
both $m$ and $a$, in the natural unit, carry the dimensions of the length.)
Then from eq.(42), it is obvious that for $m^2>3l^2_{s}$, then $m^2_{T}>0$ and for
$0<m^2<3l^2_{s}$, $m^2_{T}<0$. This result indicates that for
successful tachyon condensation in which tachyon mass squared has to become
positive definite, we should demand $m^2>3l^2_{s}$, which is consistent
with the condition for the supergravity description to be valid.
Moreover, as the ratio $(m^2/l^2_{s})$ gets larger, which certainly is in the
allowed region in the supergravity analysis, tachyon gets heavier and
eventually the tachyon, rolled down to the minimum of its potential, would
{\it decouple}, i.e., disappear from the spectrum of massless particles. 
And we identify this with a successful tachyon condensation. 
This behavior has been illustrated in Fig.2 where it can be seen
that as the ratio $(m^2/l^2_{s})$ gets larger, the potential gets deeper implying that 
the quantity representing the effective tachyon mass squared, $d^2V/d|T|^2$ evaluated
near the minimum of the potential gets bigger. 
In the mean time, for the other case $0<m^2<3l^2_{s}$, the tachyon
mass squared is still negative definite and this may signal that
although the tachyon condensation may proceed, some instability of
the system still remains. It, however, is interesting to note that
for this second case, we can, as a by-product, determine the true
vev of the tachyon field at which the minimum of the potential
occurs. That is, taking $m \simeq l_{s}$, the tachyon mass squared
becomes
\beq
m^2_{T} \simeq - {4l_{s}\over |T_{0}|^2}.
\eeq
Finally, comparing this with its genuine stringy counterpart,
$m^2_{T} = -1/\alpha'$ leads us to determine the true vev of the
open string tachyon field as
\beq
|T_{0}| = 2l^{3/2}_{s}.
\eeq
Namely, the minimum of the open string tachyon potential
$V(|T_{0}|) = -2m$ occurs when the vev of tachyon takes value of
the order $l^{3/2}_{s}$. \\
Lastly, with the supergravity analogue
of tachyon potential $V(|T|)$ at hand, one might wish to determine
the typical form of the spacetime dependent tachyon field profile
$T(x)$. We now turn to this issue. Since the tachyon arising in a
$D_{p}-\bar{D}_{p}$ system ($p=6$ in our case) is complex (and
scalar for the single $N=1$ brane-antibrane case), we write $T(x)
= |T(x)|e^{i\theta(x)}$. Then extremizing the tachyon action
\beqa
S_{T} &=& -{T_{9}\over \sqrt{\alpha'}}\int
d^{10}x[\partial_{\mu}T^{*}\partial^{\mu}T + V(|T|)]
\\ &=& -{T_{9}\over \sqrt{\alpha'}}\int
d^{10}x\sqrt{\alpha'}[\partial_{\mu}|T|\partial^{\mu}|T| +
|T|^2\partial_{\mu}\theta \partial^{\mu}\theta + V(|T|)] \nonumber
\eeqa
with respect to $|T(x)|$ and $\theta(x)$ respectively yields the
following tachyon field equations
\beqa
&&\partial_{\mu}\partial^{\mu}|T| - |T|\partial_{\mu}\theta
\partial^{\mu}\theta + m^2 l^2_{s}{1\over |T|} \left(m^2 -
l^2_{s}\ln{{|T|^2\over |T_{0}|^2}}\right)^{-3/2} = 0, \nonumber
\\ &&\partial_{\mu}(|T|^2\partial^{\mu}\theta ) =
|T|^2\partial_{\mu}\partial^{\mu}\theta + 2 |T|\partial_{\mu}|T|
\partial^{\mu}\theta = 0.
\eeqa
Now, since the true vacuum to which the tachyon condenses is
$V(|T_{0}|)=-2m$ at which $|T| = |T_{0}| = 2l^{3/2}_{s}$, we look
for a finite action solution to these field equations which
asymptotes to $|T_{0}|$ at $|x|\rightarrow \infty$. From the
second equation, it is rather obvious that the most naive solution
involves $\theta(x) = constant$ in which case the first equation
reduces to
\beq
\partial_{\mu}\partial^{\mu}|T| +
m^2 l^2_{s}{1\over |T|} \left(m^2 - l^2_{s}\ln{{|T|^2\over
|T_{0}|^2}}\right)^{-3/2} = 0
\eeq
while the next simplest solution would involve
$\partial^{\mu}\theta(x) = c^{\mu}/|T|^2$ (where $c^{\mu}$ is a
constant vector normalizable to $c^{\mu}c_{\mu}=l^4_{s}$) in which
case the first equation becomes
\beq
\partial_{\mu}\partial^{\mu}|T| - {l^4_{s}\over |T|^3}
+ m^2 l^2_{s}{1\over |T|} \left(m^2 - l^2_{s}\ln{{|T|^2\over
|T_{0}|^2}}\right)^{-3/2} = 0.
\eeq
These equations may be solved analytically or at least
numerically. We may have more to say on this in our future work.

\section{Summary and discussions}\label{ }

To summarize, in this work, using an exact supergravity solution
representing the $D6-\bar{D6}$ system, we demonstrated in a
rigorous fashion that one can construct a supergravity analogue of
the tachyon potential which may possess generic features of the
genuine stringy tachyon potentials. In doing so, our philisophy was
to evaluate the interaction energy between the brane and the
antibrane for small but finite inter-brane separation and then
identify it with the tachyon potential $V(T)$. And this
identification demands an appropriate suggestion to relate the
parameter $a$ in the supergravity solution representing the
inter-brane distance to the tachyon field expectation value and we
proposed an ansatz given in eq.(31). For the tachyon potential
thus obtained, its minimum value was of particular interest in
association with Sen's conjecture for the tachyon condensation on
unstable $D$-branes in type II theories. In the absence of the
magnetic field (i.e., $RR$ $F7$-brane) content, minimum of the
potential was unbounded below and it is due to the Coulomb-type
combined gravitational and gauge attractions between the brane and
the antibrane when they are close enough to each other. Then by
introducing the magnetic field content with an appropriate
strength $B$ removing all the conical singularities of the
$D6-\bar{D6}$ solution, the tachyon potential has been {\it
regularized}, i.e., made to be bounded below. Remarkably, now the
finite minimum value of the potential turned out to be
$V(T_{0})=-2m$. Since $m$ denotes the ADM mass of a single
$D6$-brane, this result, in a sense, appears to confirm that Sen's
conjecture is indeed correct although the analysis used is
semi-classical in nature and hence should be taken with some care.
Also commented was the fact that the supergravity analogue of the
tachyon potential constructed in the present work can only be that
in the $D_{p}-\bar{D}_{p}$ system and we need some other approach
to obtain the tachyon potential in the other non-BPS
$D_{p}$-branes, i.e., the $D_{p}$-branes with ``wrong'' $p$. We
also read off, from the vanishing separation (between $D6$ and
$\bar{D6}$) limit of the supergravity analogue of the tachyon
potential, the value of the supergravity analogue of the tachyon
mass squared $m^2_{T}$.  Then we demonstrated that the quantity
$m^2_{T}$ (which has started out as being negative) can actually
become positive definite as the tachyon rolls down toward the
minimum of its potential. It indeed signals the possibility of successful
condensation of the tachyon since it shows that near the minimum
of its potential, tachyon can become as heavy as desired within 
the validity of supergravity description and disappear from
the spectrum. Lastly, we provided tachyon field equations from
which one can determine, as a solution, the spacetime dependent tachyon field
profile $T(x)$. \\ It is also interesting to note that lately,
motivated by the series of suggestions made by Sen \cite{sen3},
some intriguing possibility of tachyonic inflation on unstable
non-BPS branes has been proposed and explored \cite{inflate} . 
There, the tachyon
field, whose dynamics on a non-BPS brane is described by an
extended version of Dirac-Born-Infeld (DBI) and Chern-Simons (CS)
action terms containing both the kinetic and potential energy
terms for the tachyon field, plays the role of ``inflaton''
driving an inflationary phase on the brane world. Clearly, the
most crucial condition for the occurrence of ``slow roll-over''
regime in a new inflation-type model is the nearly flat structure
of the tachyon potential. Tachyon potential with such highly
restricted behavior has not been available when looked for in the
context of the open (super) string field theory constructions \cite{osf}. In
this sense, it is quite promising that the supergravity analogue
of the tachyon potential obtained in the present work can serve as
a successful candidate for such a flat potential since, as we have
seen in this work, it possesses purely logarithmic dependence on
the tachyon field and hence is slowly varying. Thus it will be the
issue of one of our future works to explore the potentially
successful role played by the tachyon field arising in the
unstable $D_{p}-\bar{D}_{p}$ system and having the structure of
potential discovered in the present work. \\ After all, having a
closed form for a tachyon potential at hand invites a lot of opportunity
to explore many exciting issues in string/brane physics and its
cosmological implications. It is our hope that here we actually have made
one step further in this direction.

\section*{Acknowledgments}

This work was supported in part by grant No. R01-1999-00020 from the Korea Science and Engineerinig 
Foundation and by BK21 project in physics department at Hanyang Univ.The author would like to thank 
Sunggeun Lee for assistance to generate the figures.


\begin{thebibliography}{99}


\bibitem{non-BPS}
O.~Bergman and M.~R.~Gaberdiel, Phys.\ Lett.\ {\bf B441} (1998) 133 ;
A.~Sen, JHEP {\bf 9809} (1998) 023 ; JHEP {\bf 9810} (1998) 021 ; JHEP {\bf 9812} (1998) 021 ;
E.~Witten, JHEP {\bf 9812} (1998) 019 ; 
P.~Horava, Adv.\ Theor.\ Math.\ Phys.\ {\bf 2} (1999) 1373. 


\bibitem{harvey}

J.~A.~Harvey, P.~Horava and P.~Krauss, JHEP {\bf 0003} (2000) 021 ;
M.~R.~Gaberdiel, Class.\ Quant.\ Grav.\ {\bf 17} (2000) 3483.

\bibitem{lerda1}
A.~Lerda and R. Russo, Int.\ J.\ Mod.\ Phys.\ {\bf A15} (2000) 2000. 

\bibitem{lerda2}
M.~Frau, L.~Gallot, A.~Lerda and P.~Strigazzi, Nucl.\ Phys.\ {\bf B564} (2000) 60 ;
L.~Gallot, A.~Lerda and P.~Strigazzi, Nucl.\ Phys.\ {\bf B586} (2000) 206 ;
M.~Bertolini, P.~Di ~Vecchia, M.~Frau, A.~Lerda, R.~Marotta, and R.~Russo, 
Nucl.\ Phys.\ {\bf B590} (2000) 471.

\bibitem{halpern}
K.~Bardakci, Nucl.\ Phys.\ {\bf B68} (1974) 331 ;
K.~Bardakci and M.~B.~Halpern, Phys.\ Rev.\ {\bf D10} (1974) 4230 ;
K.~Bardakci and M.~B.~Halpern, Nucl.\ Phys.\ {\bf B96} (1975) 285 ;
K.~Bardakci, Nucl.\ Phys.\ {\bf B133} (1978) 297.

\bibitem{sen1}
A.~Sen, JHEP {\bf 9808} (1998) 010 ; JHEP {\bf 9808} (1998) 012 ; JHEP {\bf 9912} (1999) 027 ;
hep-th/9904207.

\bibitem{osf}
E.~Witten, Nucl.\ Phys.\ {\bf B268} (1986) 253 ;
V.~A.~Kostelecky and S.~Samuel, Phys.\ Lett.\ {\bf B207} (1988) 169 ; Nucl.\ Phys.\ {\bf B336} (1990) 263 ;
A,~Sen and B.~Zwiebach, JHEP {\bf 0003} (2000) 002 ;  N.~Berkovits, A.~Sen and B.~Zwiebach, Nucl.\ Phys.\ {\bf 587} (2000) 147 ;
N.~Berkovits, JHEP {\bf 0004} (2000) 022 ; N.~Moeller, A.~Sen and B.~Zwiebach, JHEP {\bf 0008} (2000) 039 ;
N.~Moeller and W. Tayler, Nucl.\ Phys.\ {\bf B583} (2000) 105 ; J.~A.~Harvey and P.~Krauss, JHEP {\bf 0004} (2000) 012 ;
R.~de ~Mello ~Koch, A.~Jevicki, M.~Mihailescu and R.~Tatar, Phys.\ Lett.\ {\bf B482} (2000) 249.

\bibitem{guijosa}
Ulf ~H.~Danielsson, A.~Guijosa and M.~Kruczenski, JHEP {\bf 0109} (2001) 011. 

\bibitem{koji}
K.~Hashimoto, JHEP {\bf 0207} (2002) 035. 

\bibitem{sen2}
A.~Sen, JHEP {\bf 9710} (1997) 002.

\bibitem{gp}
D.~Gross and M.~Perry, Nucl.\ Phys.\ {\bf B226} (1983) 29.

\bibitem{bonnor1}
W.~B.~Bonnor, Z.\ Phys.\ {\bf 190} (1966) 444 ;
S.~Mukherji, hep-th/9903012 ; B.~Janssen and S.~Mukherji, hep-th/9905153.

\bibitem{emp}
R.~Emparan, Phys.\ Rev.\ {\bf D61} (2000) 104009 ;
A.~Chattaraputi, R.~Emparan and A.~Taormina, Nucl.\ Phys.\ {\bf B573} (2000) 291.

\bibitem{bonnor2}
A.~Davidson and E.~Gedalin, Phys.\ Lett.\ {\bf B339} (1994) 304 ;
D.~V.~Galtsov, A.~A.~Garcia and O.~V.~Kechkin, Class.\ Quant.\ Grav.\ {\bf 12} (1995) 2887.

\bibitem{youm}
D.~Youm, Nucl.\ Phys.\ {\bf B573} (2000) 223.

\bibitem{ernst}
F.~J.~Ernst, J.\ Math.\ Phys.\ {\bf 17} (1976) 515.

\bibitem{bmo}
P.~Brax, G.~Mandal and Y.~Oz, Phys.\ Rev.\ {\bf D63} (2001) 064008.

\bibitem{teo}
Y.~C.~Liang and E.~Teo, Phys.\ Rev.\ {\bf D64} (2001) 024019.

\bibitem{hawking}
S.~W.~Hawking and G.~T.~Horowitz, Class.\ Quant.\ Grav.\ {\bf 13} (1996) 1487.

\bibitem{sen3}
A.~Sen, hep-th/0203211 ; hep-th/0203265 ; hep-th/0204143.

\bibitem{inflate}
G.~W.~Gibbons, hep-th/0204008 ;
M.~Fairbairn and M.~H.~G.~Tytgat, hep-th/0204070.

\end{thebibliography}
\end{document}